\documentclass[aps,pre,superscriptaddress,twocolumn,twoside,a4paper,floatfix]{revtex4}
\usepackage{hyperref,graphicx,amsmath,amssymb,mathpazo,color}
\usepackage[normalem]{ulem}

\newcommand{\tr}{\textrm{tr}}

\newcommand{\ket}[1]{|#1\rangle}
\newcommand{\Exp}[1]{\langle#1\rangle}
\newcommand{\kets}[2]{|#1\rangle_{\!_#2}}
\newcommand{\bras}[2]{{}_{_#2\!\!}\langle#1|}
\newcommand{\proj}[2]{|#1\rangle_{\!_#2\!\!}\langle#1|}

\begin{document}
\title{Virtual qubits, virtual temperatures, and the foundations of thermodynamics}

\author{Nicolas Brunner} \affiliation{H. H. Wills Physics Laboratory, University of Bristol$\text{,}$ Tyndall Avenue, Bristol, BS8 1TL, United Kingdom}
\author{Noah Linden} \affiliation{Department of Mathematics, University of Bristol$\text{,}$ University Walk, Bristol BS8 1TW, United Kingdom}
\author{Sandu Popescu} \affiliation{H. H. Wills Physics Laboratory, University of Bristol$\text{,}$ Tyndall Avenue, Bristol, BS8 1TL, United Kingdom}
\author{Paul Skrzypczyk}\affiliation{H. H. Wills Physics Laboratory, University of Bristol$\text{,}$ Tyndall Avenue, Bristol, BS8 1TL,
United Kingdom} \affiliation{Department of Applied Mathematics and Theoretical Physics$\text{,}$ University of
Cambridge, Centre for Mathematical Sciences, Wilberforce Road, Cambridge CB3 0WA, United Kingdom}
\begin{abstract}
We argue that thermal machines can be understood from the perspective of `virtual qubits' at `virtual temperatures': The relevant way to view the two heat baths which drive a thermal machine is as a composite system. Virtual qubits are two-level subsystems of this composite, and their virtual temperatures can take on any value, positive or negative. Thermal machines act upon an external system by placing it in thermal contact with a well-selected range of virtual qubits and temperatures. We demonstrate these claims by studying the smallest thermal machines. We show further that this perspective provides a powerful way to view thermodynamics, by analysing a number of phenomena. This includes approaching Carnot efficiency (where we find that all machines do so essentially by becoming equivalent to the smallest thermal machines), entropy production in irreversible machines, and a way to view work in terms of negative temperature and population inversion. Moreover we introduce the idea of "genuine" thermal machines and are led to considering the concept of ``strength'' of work.

\end{abstract}

\maketitle

\section{Introduction}
The field of quantum thermodynamics \cite{GemMicMah04,GyfBer05,AllBalNie04} has made great strides in gaining a basic understanding of phenomena at the intersection of quantum mechanics and thermodynamics. This line of research goes all the way back to the 1950's when thermodynamic analysis of lasers was investigated \cite{Ram56, ScoSch59,GeuSchSco67}. Since then there has been significant interest in many related areas, with many results on quantum thermal machines \cite{GevKos96,BenBroMei02,LinChe03,Scully03,HumLin05,SegNit06,HenMicMah06,HenMahMic07,BouTan07,QuaLiuSun07,AllHovMah10,Scully11}, finite-time thermodynamics \cite{GevKos92,GevKos94,FelGevKos96,FelKos00,FelKos03,KosFel10} and the second law \cite{PusWor78,Len78,Lloyd89,JanWocZei00,Kie04} (and references therein).

Of particular interest to this work is one recent development, namely the study of the
smallest possible self-contained thermal machines \cite{LinPopSkr10a,SkrBruLin10,LinPopSkr10b,PalKosGor01,YouMahOba09}. By self-contained we mean that no sources of external work
or other form of control are allowed; only incoherent interactions with thermal baths at various temperatures.
These smallest self-contained thermal machines are arguably the most elementary thermal machines. As such,
they are inherently simple and transparent, they offer us a view into the core of thermodynamics unobstructed by unnecessary details and complications. What we want to show here is that because of this, their study leads to a new view of thermodynamics, allowing for general conclusions to be drawn about the way thermal machines ultimately function.

At the core of this view of thermodynamics lies the notion of \emph{a virtual qubit} and its \emph{virtual temperature.} A virtual qubit is a two-level subsystem of the two baths that drive a thermal machine, when considered as a composite system. Different virtual qubits have different virtual temperatures. Using these concepts, we will show first that the smallest thermal machines -- refrigerators, heat pumps and heat engines -- all function via a simple mechanism; they place an external system in thermal contact with a well chosen virtual qubit, at a well chosen virtual temperature. More complicated thermal machines, including classical ones, use essentially the same mechanism. The only difference is that they couple the external system to many virtual qubits, at many virtual temperatures, all at the same time. 
 
We will demonstrate further the breadth of this new notion. First of all the virtual qubit provides a natural way to understand work, in terms of population inversion and negative temperatures. This in turn shows that work has an additional property, which we term ``strength''. Additionally this notion is seen to be particularly powerful in understanding efficiency, especially the Carnot limit. The smallest thermal machines will be shown to have an efficiency which is always universal (independent of model details), whilst any thermal machine which approaches the Carnot efficiency functions, essentially, as the smallest machines in the limit, utilising only a single virtual temperature. Moreover the strength of work will be seen to vanish in the Carnot limit, reinforcing the weak nature of the Carnot limit. Finally, we will see that the resources a pair of thermal baths provide is captured by the notion of virtual temperature, which in turn allows for the idea of a `genuine thermal machine' to be introduced.

The paper is organised as follows: In section \ref{s:virtual qubit} we introduce the virtual qubit and its virtual temperature and outline in section \ref{s:thermal machines} how this notion is used to understand thermal machines. In section \ref{s:smallest machines} we begin to demonstrate explicitly this idea for the smallest thermal machines, discussing first refrigerators and heat pumps in section \ref{s:fridge pump}, and then work and heat engines in section \ref{s:engine}. In section \ref{s:strength} we introduce the `strength' of work and its manifestation for both finite-dimensional and infinite-dimensional systems. We begin our discussion of efficiency in section \ref{s:universality}, where we discuss the universality of the efficiency of the smallest machines. Continuing in section \ref{s:carnot} we discuss how thermal machines approach the Carnot limit, whilst in section \ref{s:entropy} we study entropy production in irreversible machines. Finally in section \ref{s:genuine} we introduce `genuine' machines.

\section{The virtual qubit}\label{s:virtual qubit}
Suppose we have two non-interacting thermal baths at different temperatures, $T_1$ and $T_2$. The state of each
bath by itself is rather trivial, being simply a Boltzmannian distribution. However, thermal machines work by
accessing both baths. It is therefore far more relevant to look at the two baths together, as a composite
system. The structure of this system, which may seem no more complicated than its constituent parts, has in fact
a rich structure. A ``virtual qubit'' is the most elementary subspace in the Hilbert space of the composite bath,
i.e. a two-dimensional subspace. Most important are virtual qubits in which the two (basis)
states are energy eigenstates of the composite system. What makes the virtual qubit interesting is that in
general it behaves as if it has a virtual temperature $\mathcal{T}_v$, which can be very different from either
bath temperature, $T_1$ or $T_2$; in particular $\mathcal{T}_v$ can be much smaller, or larger than either, or
even negative.

\begin{figure}[t]
	\includegraphics[height=38mm]{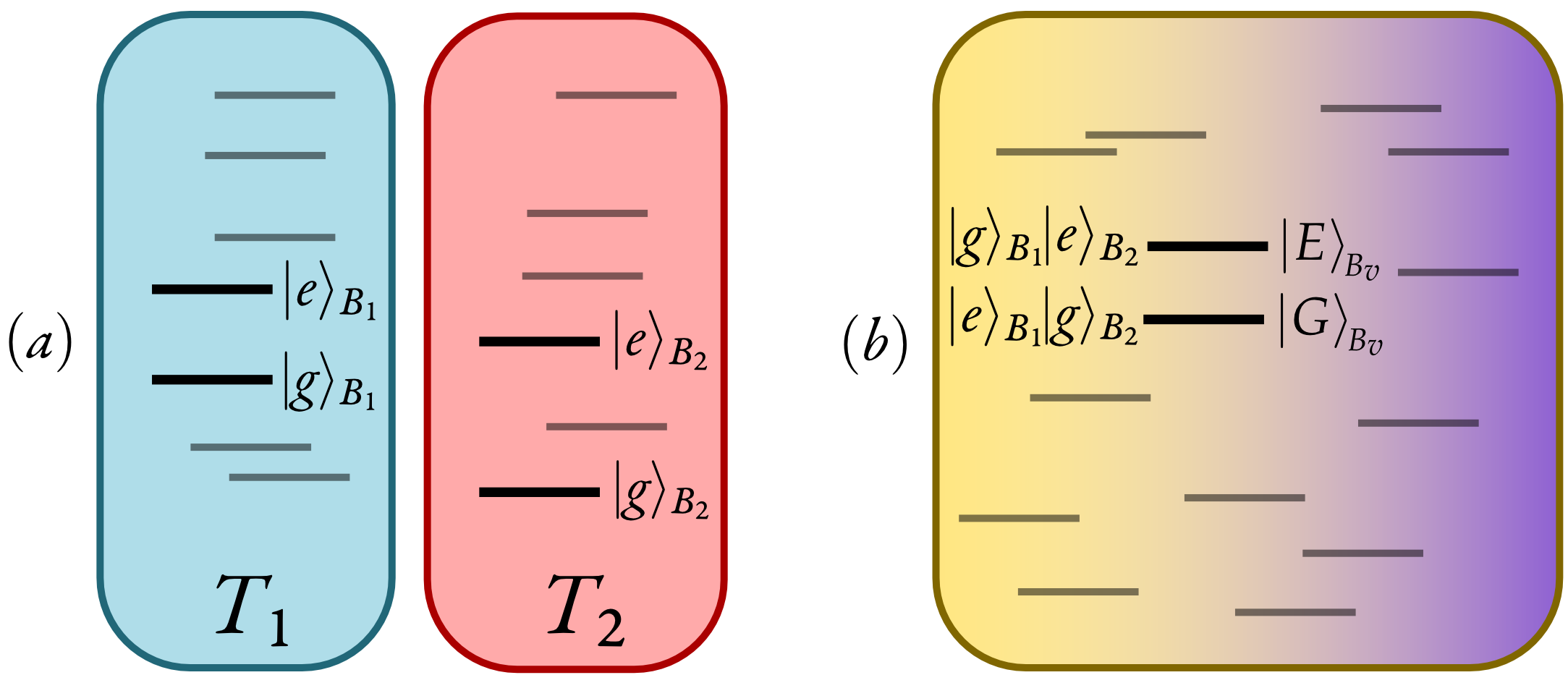} \caption{\label{f:bath-virtual} (Colour online) {\bf Schematic diagram of bath virtual qubits:} (a) Two baths, one at temperature $T_1$ and one at temperature $T_2$. In each bath we identify two arbitrary energy eigenstates, $\kets{g}{{B_i}}$ and $\kets{e}{{B_i}}$, with spacing $\mathcal{E}_i$. (b) These two baths can be viewed as a single composite bath. Two energy eigenstates of this composite system are $\kets{G}{{B_v}} \equiv \kets{e}{{B_1}}\kets{g}{{B_2}}$ and $\kets{E}{{B_v}}~\equiv~\kets{g}{{B_1}}\kets{e}{{B_2}}$. The `virtual temperature' $\mathcal{T}_v$ of this `virtual qubit' can take on any value -- positive or negative -- depending on $\mathcal{E}_1$ and $\mathcal{E}_2$. }
\end{figure}

To be more precise, let us consider two arbitrary energy eigenstates in bath 1 with energy spacing
$\mathcal{E}_1$, which we denote $\kets{g}{{B_1}}$ and $\kets{e}{{B_1}}$. Similarly, consider two states in bath 2
with energy spacing $\mathcal{E}_2$, denoted $\kets{g}{{B_2}}$ and $\kets{e}{{B_2}}$. These are depicted in FIG.~\ref{f:bath-virtual} (a). From these states we can
identify four energy eigenstates of the composite system, namely $\kets{g}{{B_1}}\kets{g}{{B_2}}$,
$\kets{e}{{B_1}}\kets{g}{{B_2}}$, $\kets{g}{{B_1}}\kets{e}{{B_2}}$ and $\kets{e}{{B_1}}\kets{e}{{B_2}}$. Of particular interest are virtual qubits formed of the states $\kets{e}{{B_1}}\kets{g}{{B_2}}$ and
$\kets{g}{{B_1}}\kets{e}{{B_2}}$. Without loss of generality we may take $\mathcal{E}_2 > \mathcal{E}_1$. With this
choice, these two states form a virtual qubit with energy spacing $\mathcal{E}_v = \mathcal{E}_2-\mathcal{E}_1$.
It is convenient to define the virtual ground and excited states as, respectively
\begin{align}
    \kets{G}{{B_v}} &=\kets{e}{{B_1}}\kets{g}{{B_2}}& \kets{E}{{B_v}} &=\kets{g}{{B_1}}\kets{e}{{B_2}}
\end{align}
depicted in FIG.~\ref{f:bath-virtual} (b). The ``virtual temperature'' of this qubit can be found by looking at the ratio of populations $p^E_v$ and
$p^G_v$, of the excited and ground state,
\begin{equation}\label{e:virtual temperature}
    e^{-\mathcal{E}_v/\mathcal{T}_v} = \frac{p^E_v}{p^G_v}
\end{equation}
Since each bath is individually in a Boltzmannian distribution, we know that the populations of the eigenstates from a single bath satisfy the relations
\begin{align}
    p^e_1 &= p^g_1 e^{-\mathcal{E}_1/T_1}& p^e_2 &= p^g_2 e^{-\mathcal{E}_2/T_2}
\end{align}
Therefore we see that
\begin{align}
    p^E_v &= p^g_1p^e_2 = p^g_1p^g_2 e^{-\mathcal{E}_2/T_2}\nonumber \\
    p^G_v &= p^e_1p^g_2 = p^g_1p^g_2 e^{-\mathcal{E}_1/T_1}
\end{align}
and so we obtain
\begin{equation}
    e^{-\mathcal{E}_v/\mathcal{T}_v} = \frac{e^{-\mathcal{E}_2/T_2}}{e^{-\mathcal{E}_1/T_1}}
\end{equation}
which allows us finally to arrive at
\begin{equation}
    \label{e:Tv bath} \mathcal{T}_v = \frac{\mathcal{E}_2-\mathcal{E}_1}{\frac{\mathcal{E}_2}{T_2}-\frac{\mathcal{E}_1}{T_1}}.
\end{equation}
as the virtual temperature of the virtual qubit. 

There are a couple of important observations which we must now make.

The first is that the virtual temperature does not depend on the energy of each individual level, but only on their energy difference.

Second, the virtual temperature, as a function of the local energy level spacings $\mathcal{E}_1$ and $\mathcal{E}_2$, can take a range of values, crucially \emph{all} temperatures outside
the range of $T_1$ and $T_2$ -- it can be smaller or larger than both, and can even take negative values \footnote{Note
that if we had instead considered virtual qubits formed of the states $\kets{g}{{B_1}}\kets{g}{{B_2}}$ and
$\kets{e}{{B_1}}\kets{e}{{B_2}}$ then the virtual temperatures would have been limited to lie between $T_1$ and $T_2$.}.

Third, in macroscopic baths we can find essentially any possible energy levels. 
Hence the composite system of a pair of macroscopic baths contains virtual qubits at essentially any virtual temperature $\mathcal{T}_v$. 

Finally, for every given virtual temperature $\mathcal{T}_v$, one can find essentially infintely many virtual qubits in the composite system. This follows from the fact that each bath locally has infinitely many qubits with the spacings $\mathcal{E}_1$ and $\mathcal{E}_2$.

It is precisely these these facts that give the composite system its thermodynamic power, as we shall show in the rest of the paper.

\section{Thermal machines}\label{s:thermal machines}
The central idea of this paper is that: 
\begin{quote}\emph{
All a thermal machine does is to place an external system in direct thermal contact with a restricted set of virtual qubits in the composite bath, having only a restricted range of virtual temperatures. The external system simply reacts to the virtual qubits as it would react if put in thermal contact with real qubits having the same virtual temperatures.}
\end{quote} 
If the temperatures are predominantly smaller than those of the individual baths,
the machine is a refrigerator. If the temperatures are predominantly larger, then it is a heat pump. Finally, if
the temperature is negative, then the machine is a heat engine. In this sense thermal machines act
simultaneously as a coupler and a filter -- they provide a coupling between the system and the thermal
reservoirs, but also filter out only a restricted range of virtual temperatures.

The important question is how exactly does a thermal machine achieve this task. In particular, how can a thermal
machine filter out selected virtual qubits from the composite bath? This ability seems surprising since we can
construct thermal machines without detailed knowledge of the spectrum of the bath. In the next section we will
give an answer to this question for the case of the smallest thermal machines. In fact, as we will see, the
smallest thermal machines couple not to a range of virtual temperatures, but to only a \emph{single} one. In
this sense, they act as perfect filters and are therefore the ``cleanest'' thermal machines. We will see later in Section \ref{s:carnot} that the ability to perfectly filter is intimately related to the ability to reach Carnot efficiency.

\section{How thermal machines work: The smallest thermal machines}\label{s:smallest machines}
\subsection{The machine virtual qubit}
\begin{figure}
	\includegraphics[height=86mm]{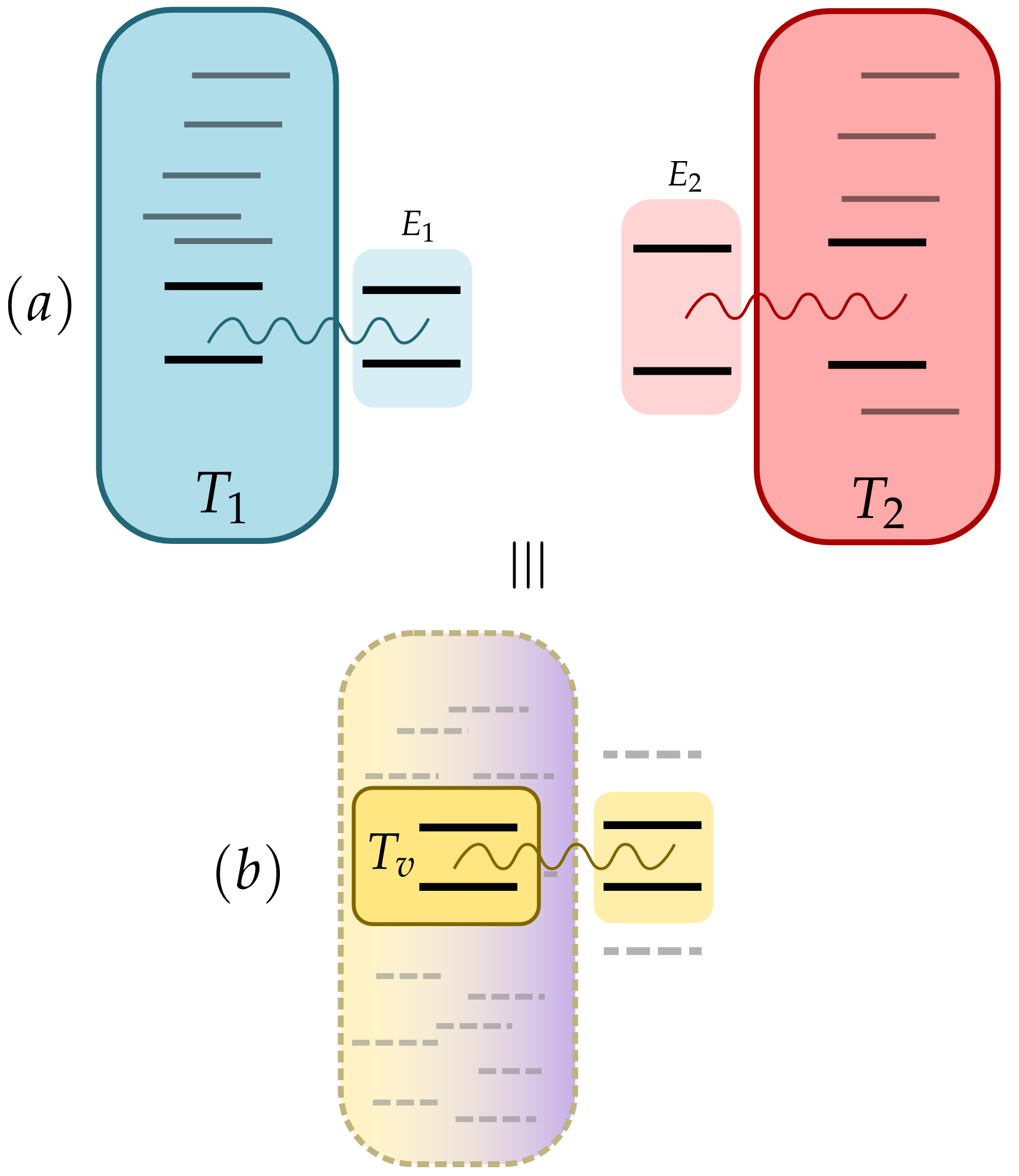} \caption{\label{f:small-machine-equiv} (Colour online) {\bf Schematic diagram of the smallest thermal machines:} (a) The smallest machine comprises two qubits, one with energy spacing $E_1$ in thermal contact (wavy line) with a bath at temperature $T_1$ and the other with energy spacing $E_2$ in thermal contact with a bath at temperature $T_2$. (b) The ultimately more relevant way to view this system is that the `machine virtual qubit' with spacing $E_v = E_2-E_1$, is in thermal contact with virtual qubits in the composite bath at the virtual temperature $T_v$. The local energy-level spacings $E_1$ and $E_2$ allow the machine virtual qubit to `filter' out a single virtual temperature from the composite bath.}
\end{figure}

All of the smallest thermal machines have at their core the same basic mechanism. Each machine consists of two
qubits, qubit 1 that is in thermal contact with bath 1, and qubit 2 that is in thermal contact with bath 2 (See FIG.~\ref{f:small-machine-equiv} (a)).
These two qubits will then interact with each other and an external system via an interaction Hamiltonian. It is
illuminating to consider the state of the two qubits before this interaction is turned on. Suppose that qubit 1
has energy spacing $E_1$ and qubit 2 energy spacing $E_2$. The explicit choice of these energies is part of the
design of the machine and is chosen by us. In the absence of interaction with the external system, each qubit
interacts only with its own thermal bath. As such, each will reach thermal equilibrium at the corresponding bath
temperature.

For simplicity, throughout this paper we will only consider the case of weak coupling between the machine and
the bath, as this will allow us to focus on \emph{resonant} (i.e. energy conserving) interactions only. In this regime the thermal state of each qubit will be a Boltzmannian distribution with the
Hamiltonian being the free Hamiltonian of the qubit. Thus the thermal state of each qubit is given by
\begin{eqnarray}
    \tau_i &=& \tfrac{1}{\mathcal{Z}_i}e^{-H_i/T_i} \nonumber \\
    &=& \frac{1}{1+e^{-E_i/T_i}}\Big(\proj{0}{i} + e^{-E_i/T_i}\proj{1}{i}\Big)
\end{eqnarray}
where $H_i = E_i \proj{1}{i}$ is the free Hamiltonian of each qubit.

Let us look at this trivial thermalisation process, in which each qubit simply reaches equilibrium with its own
bath, from a different, and ultimately more relevant, point of view. The two qubits have 4 energy eigenstates,
$\kets{0}{1}\kets{0}{2}$, $\kets{0}{1}\kets{1}{2}$, $\kets{1}{1}\kets{0}{2}$ and $\kets{1}{1}\kets{1}{2}$. Of particular interest are the two states
$\kets{0}{1}\kets{1}{2}$ and $\kets{1}{1}\kets{0}{2}$. In similar fashion to the previous section, let us interpret the Hilbert space spanned
by these two states as a virtual qubit, with ground and excited state
\begin{align}\label{e:machine virtual}
\kets{0}{v} &= \kets{1}{1}\kets{0}{2}& \kets{1}{v} &= \kets{0}{1}\kets{1}{2}
\end{align}
This will be our \emph{machine virtual qubit}. The energy spacing of these two states is $E_v = E_2 = E_1$. By
repeating a similar analysis to that carried out for the virtual qubit in the composite bath, we find that this
machine virtual qubit reaches the temperature $T_v$
\begin{equation}
    \label{e:Tv machine} T_v = \frac{E_2-E_1}{\frac{E_2}{T_2}-\frac{E_1}{T_1}}.
\end{equation}
We can interpret this result as saying that \emph{the machine virtual qubit entered into thermal contact with
virtual qubits in the composite bath and thermalised to their temperature}.

Seen from this point of view, the thermalisation process may seem surprising. As we saw previously, there are
many virtual qubits in the composite bath having the same energy-level spacing as the machine virtual qubit, i.e. satisfying $\mathcal{E}_2-\mathcal{E}_1=E_2-E_1$. It would appear therefore that all such virtual qubits should couple resonantly to the machine virtual qubit. However, the temperature of the virtual qubits depends not only on the level spacing $\mathcal{E}_2-\mathcal{E}_1$, but also on the separate values of $\mathcal{E}_2$ and $\mathcal{E}_1$; coupling all of these to the machine virtual qubit will result in it attaining some average temperature. The key to understanding the behaviour of the machine virtual qubit is to realise that the interaction with the
composite bath is not arbitrary; it occurs only via the \emph{local} couplings of qubit 1 to bath 1 and qubit 2
to bath 2. Each machine qubit will only interact resonantly with a corresponding qubit in the bath, so $\mathcal{E}_1=E_1$ and $\mathcal{E}_2=E_2$.
This allows the machine to couple its virtual qubit in a very selective way to the virtual qubits in the
composite bath and to single out the precise temperature $T_v$. In the terminology of the previous section, we
see that it is the local resonant couplings which provide the filtering for these machines. Furthermore, we see that only a single $T_v$ is selected, and as such this machine acts as a perfect filter. Finally, we note that the specific value of $T_v$ filtered out is chosen at our disposal by the the way in which we engineer the machine, i.e. by the choice of $E_1$ and $E_2$.

\subsection{Coupling to the external system}
 We now have all the necessary tools to understand the behaviour of these thermal machines: \emph{All we have to do
 now is to place the external system in thermal contact with the virtual qubit of the machine.}

As an example, let us consider the simplest case, where the system which the machine will act upon is itself
another qubit. This qubit shall be called qubit 3 and shall have energy spacing $E_3$. Given this spacing, we
ensure that the machine is engineered such that $E_2-E_1 = E_3$. With this choice, we see that the machine
virtual qubit and the external system have equal energy-level spacings. It is therefore possible to introduce an
arbitrarily weak interaction which allows the virtual qubit and external system to resonantly exchange energy --
i.e. we place them in thermal contact.

This can be done via the interaction Hamiltonian
\begin{equation}
    \label{e:Hint} H_{int} = g\big(\kets{0}{1}\kets{1}{2}\kets{0}{3}\bras{1}{1}\bras{0}{2}\bras{1}{3} + \kets{1}{1}\kets{0}{2}\kets{1}{3}\bras{0}{1}\bras{1}{2}\bras{0}{3}\big)
\end{equation}
which was previously introduced in \cite{LinPopSkr10a} and induces transitions between the two degenerate states
$\kets{0}{1}\kets{1}{2}\kets{0}{3} \leftrightarrow \kets{1}{1}\kets{0}{2}\kets{1}{3}$.  Again the most relevant way to write this Hamiltonian is in terms of the virtual qubit,
\begin{equation}
    H_{int} = g\big(\kets{0}{v}\kets{1}{3}\bras{1}{v}\bras{0}{v} + \kets{1}{v}\kets{0}{3}\bras{0}{v}\bras{1}{v}\big)
\end{equation}
which induces transitions between the two degenerate states $\kets{0}{v}\kets{1}{3} \leftrightarrow
\kets{1}{v}\kets{0}{3}$. In this notation is becomes manifest that the interaction Hamiltonian is the one which
generates the unitary {\sc swap} between the external system and the machine virtual qubit. Hence this
interaction places these two qubits into thermal contact.

\subsection{Summary}
We now reach the crux of our analysis. We see that the way in which the whole process works is that selected
virtual qubits of the composite bath, at a selected virtual temperature, are placed in thermal contact with a
virtual qubit of the machine. In turn this virtual qubit of the machine is placed in thermal contact with the
external system. Hence all that happens is an ordinary thermalisation process of the external system with the
virtual qubits in the composite bath, mediated by the machine virtual qubit. If the external system is
otherwise isolated from an external environment, then it reaches the temperature $T_v$ (for an explicit calculation see Appendix \ref{a:bounded}, and FIG.~\ref{f:small-machine}). On the other hand, if
the system is also in contact with an external environment, then it will reach some (stationary) state similarly
to any system in contact with two thermal baths, see FIG.~\ref{f:small-machine-bath}
\begin{figure}
	\includegraphics[height=40mm]{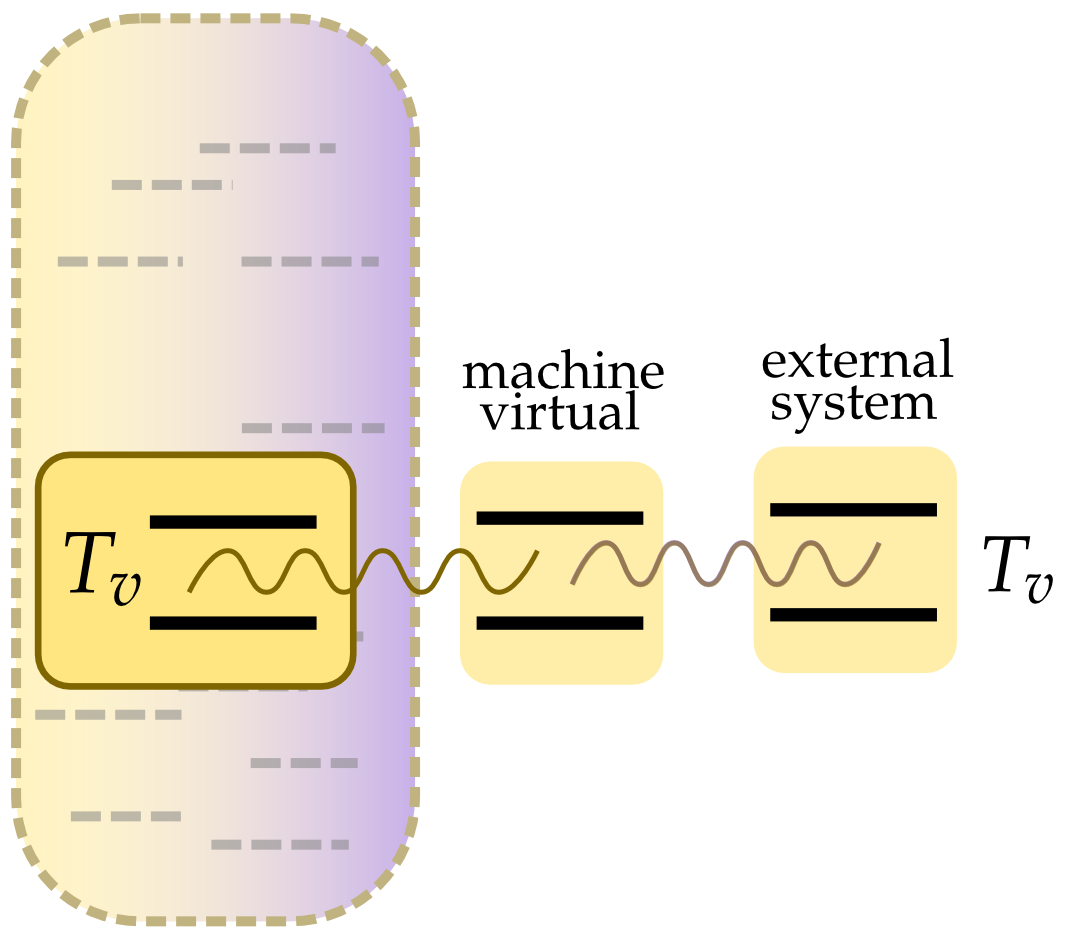} \caption{\label{f:small-machine} (Colour online) {\bf Schematic diagram of smallest machine interacting with an isolated external qubit:} An isolated external system -- here a single qubit -- is placed into thermal contact with the machine virtual qubit, which has matching energy-level spacing. This interaction is depicted by the wavy line. The net effect is that the external system is placed in thermal contact with virtual qubits in the composite bath at temperature $T_v$, mediated by the machine virtual qubit. }
	\includegraphics[height=40mm]{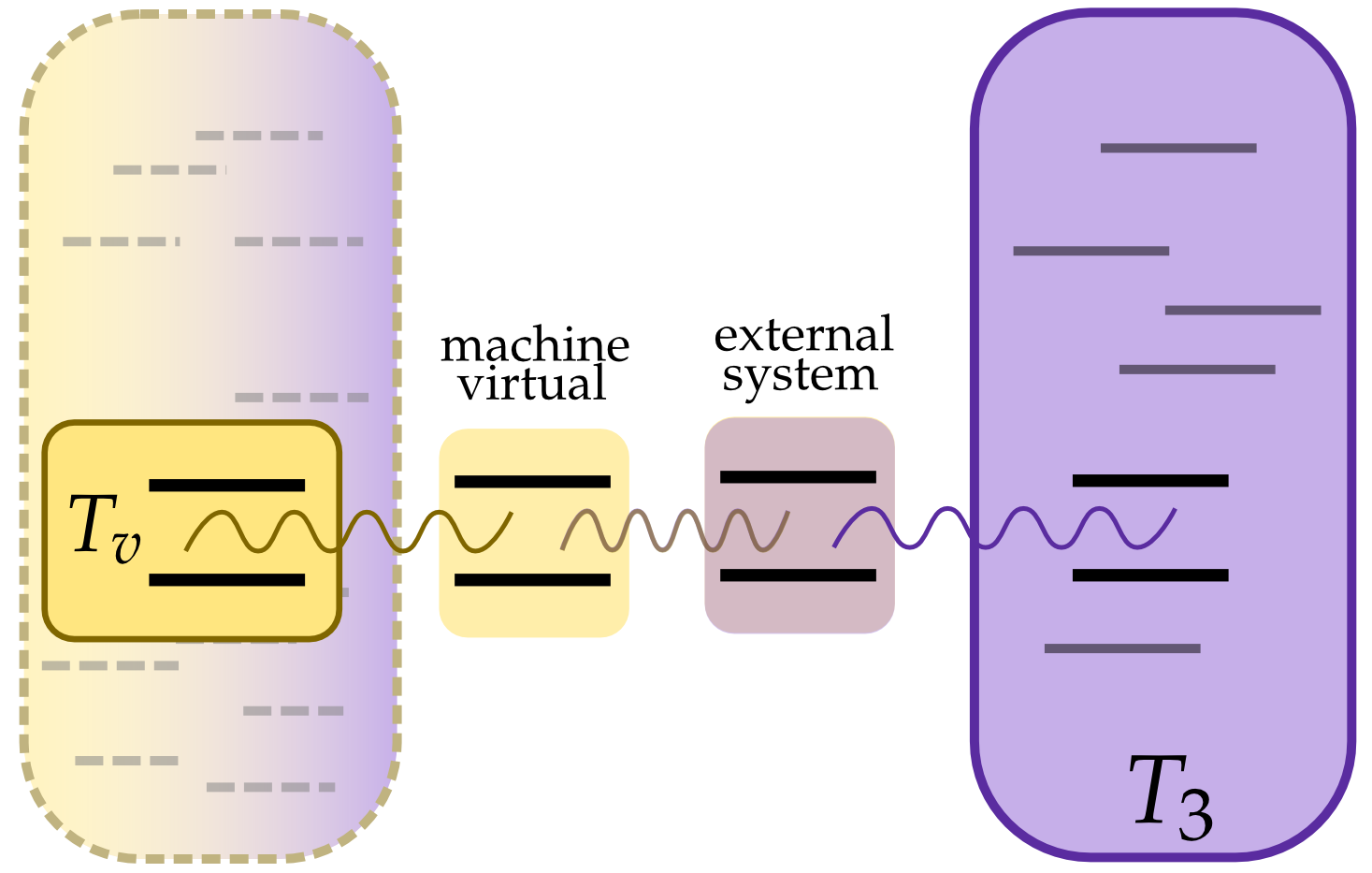} \caption{\label{f:small-machine-bath} (Colour online) {\bf Schematic diagram of smallest machine interacting with an open external qubit:} If the external system also interacts with its own environment at temperature $T_3$ then it will behave as any system in contact with two thermal baths at two differing temperatures.}
\end{figure}

We conjecture that this process is in fact a special case of the general procedure which all thermal machines
use. However, a thermal machine generally has many different energy levels, hence many different machine virtual 
qubits. In turn they couple to many different virtual qubits in the composite bath, covering a range of
virtual temperatures, instead of a single one, thus complicating the situation. It is also further complicated by the fact that when the
interaction is stronger the energy levels of the machine become broadened, so even a single virtual qubit in the
machine can couple to many virtual temperatures in the composite bath. Nevertheless, the principle remains the
same; the external system is placed in thermal contact with the various virtual qubits in the bath, mediated
by the virtual qubits of the machine.

\section{Refrigerators, heat pumps and more}\label{s:fridge pump}
As we already noted above, the particular function that the thermal machine provides depends upon the range of
virtual temperatures of the composite bath which are coupled to the external system. If these are predominately
lower than $T_1$ and $T_2$ (the actual temperatures of the two baths) then the machine is functioning as a
refrigerator. On the other hand, if the range of virtual temperatures are predominately higher than $T_1$ and
$T_2$ then the machine is functioning as a heat pump. 

A more interesting situation is what happens when the range of virtual temperatures is predominantly negative.
This corresponds to virtual qubits having population inversions. The thermodynamic significance of this and related effects is discussed in the next section.

\section{Work and Heat Engines}\label{s:engine}
\begin{figure*}
	\includegraphics[height=40mm]{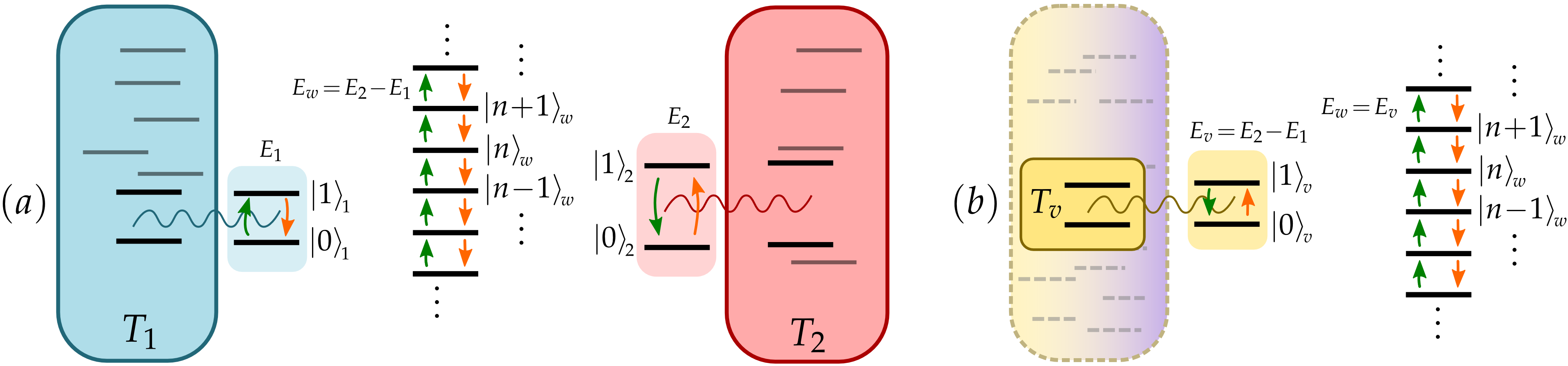} \caption{\label{f:small-engine} (Colour online) {\bf Schematic diagram of the smallest heat engine:} (a) The weight -- an equi-spaced system, unbounded from above and below, interacts with the machine qubits via the interaction depicted by vertical arrows. This interaction induced transitions between the degenerate energy eigenstates $\kets{0}{1}\kets{1}{2}\kets{n}{w} \leftrightarrow \kets{1}{1}\kets{0}{2}\kets{n\!+\!1}{w}$. The energies $E_1$ and $E_2$ are chosen such that the `forward' transition in which the weight is lifted is biased in favour of the `backward' transition in which the weight is lowered. As such the weight is lifted on average, and hence this system produces work. (b) From the viewpoint of the virtual qubit, the biasing condition says that the virtual qubit must have a population inversion, or in other words must be at a negative temperature.}
\end{figure*}

Work is one of the central concepts of physics. In comparison, population inversion is an interesting
phenomenon, but until now usually associated mostly with laser physics. In this section however, we argue
that inversion is a far more important thing -- in fact we argue that
\begin{quote}
\emph{Producing work and generating population inversion are one and the same thing}.
\end{quote}
Work is generally associated with ordered movement. Therefore the first requirement in order to be able to talk about work is to have a system with many states, otherwise the whole idea of the system having an ordered movement i.e. going from one state to another in a
systematic way, makes no sense. Talking therefore of work when dealing with systems with a small, limited,
number of states, such as when considering the two qubits comprising the thermal machine presented in the
previous section, may seem impossible. To tackle this problem many alternative ways of defining work have been
given in the literature \cite{Ali79,BouTan06a,BouTan06b,WeiHenRem08}, each with their own merits. Here we take the most direct route: just as Carnot said {\it ``motive power (work) is the useful effect that a motor is capable of
producing. This effect can always be likened to the elevation of a weight to a certain height.'' }\cite{Car24}.
Hence we consider lifting a weight to be producing work, and any machine that lifts a weight to be a machine that produces work. Following this idea we will argue that work is nothing other than producing population inversion.

At a very elementary level, this definition is clearly consistent with the usual one. Indeed, consider a weight
taken initially to be on the floor -- this is the state of lowest energy and is therefore its ground state.
Lifting the weight -- that is, doing work on it -- is nothing but completely depopulating the ground state and
populating an excited state, a state where the weight is higher up. Thus the final state of the weight is an
inverted state. Alternatively, we could consider a free particle whose kinetic energy is steadily increased by the machine, or indeed any other similar system. There is however far more here than meets the eye.

To begin with we introduce the model of the smallest heat engine \cite{LinPopSkr10a,YouMahOba09}. It is identical to that of the refrigerator,
(i.e. we take 2 qubits, one connected to a bath at $T_1$ and one connected to a bath at $T_2$). We imagine that
the engine delivers work by pulling up a weight. The weight is isolated from both baths. To simplify the
situation we consider that the weight is pulled up very slowly, so that we can neglect the change of its kinetic
energy and consider only the potential energy. Furthermore, we suppose that the weight can be situated only at
some discrete equidistant heights, so that the energy difference between them is the same. Hence, the weight has
discrete energy eigenstates $\ket{n}_w$ with corresponding energy eigenvalues $E_n = n E_w$, with $E_w > 0$.
Alternatively, we can imagine that the engine delivers work by pumping energy into a harmonic oscillator; both
situations are formally almost equivalent (the harmonic oscillator energies being limited from below). This set-up is depicted diagrammatically in FIG.~\ref{f:small-engine} (a).

The free Hamiltonian of the system is thus
\begin{equation}\label{e:free Hamiltonian}
    H_0=E_1\proj{1}{1}+E_2\proj{1}{2} + \sum_{n=-\infty}^{\infty} n E_w\proj{n}{w}
\end{equation}
The energies are taken such that
\begin{equation}
    E_2 - E_1 = E_w.\label{energy_conservation}
\end{equation}
Given this constraint, the energy levels $\kets{0}{1}\kets{1}{2}\kets{n}{w}$ and $\kets{1}{1}\kets{0}{2}\kets{n\!+\!1}{w}$ are
degenerate. The engine acts by making transitions between these degenerate states.

The qubits and weight interact via the Hamiltonian
\begin{multline}
    H_{int}= g\sum_{n=-\infty}^{\infty}\kets{0}{1}\kets{1}{2}\kets{n}{w}\bras{1}{1}\bras{0}{2}\bras{n\!+\!1}{w} \\ + \kets{1}{1}\kets{0}{2}\kets{n\!+\!1}{w}\bras{0}{1}\bras{1}{2}\bras{n}{w}
\end{multline}

The intuitive idea behind the design of this engine is to bias the transition
\begin{equation}
\kets{0}{1}\kets{1}{2}\kets{n}{w}\rightarrow\kets{1}{1}\kets{0}{2}\kets{n\!+\!1}{w}
\end{equation}
in which the weight is lifted in favour of the reverse transition  in which the weight is lowered. This is
obtained by coupling the two qubits to heat baths at different temperatures, $T_2> T_1$, chosen such that the
probability for the qubits to be initially in the state $\kets{0}{1}\kets{1}{2}$ is larger than the probability to
be in the state $\kets{1}{1}\kets{0}{2}$.

In terms of the virtual qubit of the machine \eqref{e:machine virtual} what we want is to bias the transition

\begin{equation}
\kets{1}{v}\kets{n}{w}\rightarrow\kets{0}{v}\kets{n+1}{w}
\end{equation}
in favour of the reverse transition (see FIG.~\ref{f:small-engine} (b)). The condition for this to happen is simply that
the probability to be in the state $\kets{1}{v}$ should
be larger than the probability to be in the state $\kets{0}{v}$.

\emph{In other words, the condition for the
machine to work as a heat engine is precisely that the machine virtual qubit has a population inversion and is
therefore at a negative virtual temperature.}

\subsection{Engine Model details}
So far our conclusions have been of a general nature; while they are enough to understand qualitatively the working of our heat engine it is illuminating to understand it also more quantitatively. To do this we need to say more about the way in which the two machine qubits interact with their respective heat baths. In Appendix \ref{a:heat engine} we consider an explicit model for the thermalisation of the machine qubits, identical to that previously presented for the refrigerator \cite{LinPopSkr10a}. This allows us to analytically solve for the time evolution of the system. We shall present here the important aspects of the solution.

To begin with we must first introduce two new quantities relating to the virtual qubit. The first is the equilibrium ``bias'' of the virtual qubit,
\begin{align}
	\Exp{Z_v^{eq}} &= \bras{0}{v}\tau_1\otimes\tau_2\kets{0}{v} - \bras{1}{v}\tau_1\otimes\tau_2\kets{1}{v} \nonumber \\
	&\equiv \bras{1}{1}\bras{0}{2}\tau_1\otimes\tau_2\kets{1}{1}\kets{0}{2} - \bras{0}{1}\bras{1}{2}\tau_1\otimes\tau_2\kets{0}{1}\kets{1}{2} \nonumber \\
	&\propto 1 - e^{-E_v/T_v}
\end{align}
which gives the difference in population of the ground and excited state of the virtual qubit at thermal equilibrium (i.e. in the absence of interaction with the external system). Thus whenever there is more population in the virtual ground state then the bias is positive and so too is the virtual temperature. However, when we have a population inversion the bias and virtual temperature become negative.

Second, we will need the equilibrium normalisation of the virtual qubit, 
\begin{align}
	\Exp{N_v^{eq}} &= \bras{0}{v}\tau_1\otimes\tau_2\kets{0}{v} + \bras{1}{v}\tau_1\otimes\tau_2\kets{1}{v} \nonumber \\
	&\equiv \bras{1}{1}\bras{0}{2}\tau_1\otimes\tau_2\kets{1}{1}\kets{0}{2} + \bras{0}{1}\bras{1}{2}\tau_1\otimes\tau_2\kets{0}{1}\kets{1}{2} \nonumber \\
	&\propto 1 + e^{-E_v/T_v}
\end{align}
which is simply the combined population in the virtual ground and excited states. Having introduced these two quantities we can now state the important features of the solution. To see that the weight is being raised in time, we must look at the average energy of the weight, $\Exp{E_w}$. Asymptotically we find that, from \eqref{e:rate}
\begin{equation}\label{e:Ew}
\Exp{E_w} = -\alpha E_w \Exp{Z_v^{eq}}t
\end{equation}
where $\alpha = \frac{g^2p}{2g^2+p^2}$ is a positive constant which depends only upon the strength of the interaction Hamiltonian ($g$) and the strength of the thermalisation ($p$). We thus see that the energy of the weight is exactly proportional to the bias of the virtual qubit $\Exp{Z_v^{eq}}$, and thus increases in time whenever the virtual qubit is at a negative temperature, confirming our claim that the ``weight is lifted''.

Notably, we find that the weight is not only lifted in time, but that it spreads also. The expression obtained for the spread $\Delta E_w^2$ in the asymptotic limit is, from \eqref{e:rate sq}
\begin{equation}\label{e:Delta Ew}
\Delta E_w^2 = E_w^2\big(\alpha \Exp{N_v^{eq}} - \beta \Exp{Z_v^{eq}}^2\big)t
\end{equation}
where $\beta = \frac{2g^4p(g^2+2p^2)}{(2g^2+p^2)^3}$ is a second, model-dependent positive constant.

Given the above, we see that the behaviour of the weight can be likened to that of a \emph{biased random walk}: In this picture the virtual qubit plays the role of the coin, with the `normalised bias' $\Exp{Z_v^{eq}}/\Exp{N_v^{eq}}$ playing the role of the (average) bias of the coin, and $\Exp{N_v^{eq}}$ modulating, along with $g$ and $p$, the probability per unit time to make a transition.

\section{Strength of work}\label{s:strength}

Classically when one thinks of work, one thinks of work full stop -- there is no notion of there being different
types of work. However, as we have seen above, all that is required from a heat engine in order to produce work
is that its virtual qubit should be at a negative temperature. But different heat engines may have their virtual
qubits at different negative temperature. For conciseness we will refer to this as to the "temperature at which
the work is delivered" or the "temperature of the work".  It is quite clear that work delivered at different
temperatures must be in many physical ways different. In fact, intuitively, work delivered at negative temperatures with smaller absolute values -- corresponding to larger inversions in the machine -- is in some sense ``stronger''.

It is important to emphasise that this "strength" of work is not equivalent to power, i.e. with the rate at
which work is delivered; it is a completely different notion and has nothing to do with time scales. True, as we
have seen in \eqref{e:Ew} everything else being the same, the power of a machine depends on the temperature of
the work - the smaller in absolute value the negative temperature, the larger the power. But power depends upon many other factors
besides this temperature; critically, it depends on the coupling constants which are the ones that encode the
time scales. On the other hand the virtual temperature at which work is delivered is a notion independent of any
time considerations.

The question that arises is how does the difference in virtual temperature of the work manifest itself
physically, apart from its influence on power? We do not yet know all of the ways in which it does, but there
are two important scenarios which we shall demonstrate below.

\subsection{Producing inversions in systems bounded from above}
In the previous section we saw that the effect of a heat engine is to steadily increase the energy of the system on which it acts. This, of course, can only happen if the system's energy spectrum is unbounded from above. But suppose that the energy spectrum has an upper bound; what happens then? Intuitively the answer is immediate: As we argued, all that a thermal machine does is to put the external system in simple thermal contact with the machine virtual qubit (which in turn is in thermal contact with a selected virtual qubit in the composite bath). If the external system is otherwise isolated it will tend to reach thermal equilibrium with the machine virtual qubit. If the machine virtual qubit is at negative temperature, this will be the final temperature of the external system. In a system with energy unbounded from above, such an equilibrium cannot be achieved -- in the evolution towards equilibrium, the system will forever increase its energy. On the other hand, when the energy spectrum is bounded from above, the system will reach equilibrium, thermalising to an inverse Boltzmannian distribution at the negative temperature $T_v$ (see Appendix \ref{a:bounded}).

In other words, if the weight is lifted by a string passing through a hole in a ceiling, the weight cannot raise more and will eventually reach a Boltzmannian distribution, with the maximum probability at the ceiling, but having non-zero probability to be found at all smaller heights.

Here we see one other aspect of the ``strength'' of work: The smaller in absolute value the negative temperature at which work is delivered, i.e. the ``stronger'' the work is, the closer to the ceiling the weight is pushed.

The mechanism by which the inverted Boltzmannian distribution gets established is interesting by itself. To better see
what happens consider an external system with many energy eigenstates. Suppose the system starts at an energy level far
from the top. At the initial moment the top plays no role, and the engine simply pushes it up the energy ladder. Had this evolution been uniform, without
spreading, the system would have simply ended "stuck at the ceiling", i.e. at the top energy level. But as it
climbs, the position of the weight also spreads - as stated previously, the evolution resembles closely that of a biased random walk. If initially the energy was perfectly well
defined, during the evolution it will spread approximately as a gaussian packet. This is still different from
the inverted Boltzmanian that is the final equilibrium state. However, as the system reaches the top energy level, it cannot climb in energy any further. At this moment, all it can do is to spread. As such it starts to extend
backwards, toward lower energies, reaching eventually the inverted Boltzmanian equilibrium state.

Note that the spreading, that seemed to be more like a minor side effect when studying the evolution of a system with infinite energy spectrum (or when far from the top in the case of a bounded energy one), plays in fact an essential role -- the inverted Boltzmanian equilibrium could not be achieved without it.

It is important to note also that the spreading is not restricted to quantum heat engines; classical thermal
machines lead to a spread in energy as well. Indeed, although usually one doesn't think of fluctuations in a
thermal machine, they always exist. The thermal baths that drive the machines always have fluctuations and these
lead to fluctuations in the evolution of the machine. (A nice and easy example is given by the famous Feynman
ratchet-and-pawl heat engine \cite{FeyLeiSan63}. When the pawl is up the ratchet has some probability to move backwards.)

\subsection{Energy gain versus energy spreading}

Consider again a heat engine acting on system with an infinite number of energy levels and suppose that the
system starts wit a well defined given initial energy. As we discussed before, it's evolution (in energy) is
similar to a biased random walk: as its average energy increases, it also spreads in energy. An interesting
problem is to compare the increase in average energy with the spreading.

The increase in average energy is proportional to time, while the increase in spread is proportional to the
square root of time (see equations \eqref{e:Ew} and \eqref{e:Delta Ew}). Thus for short times it is the spreading which dominates, so the system is likely to also lose energy, not only to gain it. The probability to find the system with lower energy therefore increases, even though the average energy becomes larger. After a longer time however, the gain in the average energy becomes larger than the spread.  The interesting question is not after what time the average energy becomes larger than the spread, but how much energy must have been put into the system by this time. In other words, to measure the spreading versus the scale of average energy (as opposed to in time), which is an intrinsic property of the evolution.

There is a ``break even'' time $t_{be}$ such that $\Exp{E_w}(t_{be})~=~\Delta E_w(t_{be})$. We denote the average energy at this time (the ``break even energy'') by $\Exp{E_w^{be}}$. From \eqref{e:Ew} and \eqref{e:Delta Ew} we find
\begin{equation}
	\Exp{E_w^{be}} = -E_w \Bigg(\frac{\Exp{N_v^{eq}}}{\Exp{Z_v^{eq}}} - \frac{\beta}{\alpha}\Exp{Z_v^{eq}}\Bigg)
\end{equation}
The first term in the brackets depends only upon the virtual temperature, and not on the other details of the machine. Indeed, it is easy to show that $\Exp{N_v^{eq}}/\Exp{Z_v^{eq}}~=~1/\tanh(E_v/2T_v)$. The second term however is a model-dependent term, and thus the break even energy is a non-universal property of a thermal machine. 

Crucially however, when the virtual temperature approaches minus infinity, i.e. the inversion of the machine virtual qubit becomes vanishingly small, the second term becomes negligible and the break even energy becomes universal and infinite. This is another manifestation of the strength of the work. Namely, when the strength is very weak, the break even energy becomes very large, and in the limit unobtainable. The machine therefore effectively stops working. As we will see in the next section, this is exactly the moment when the machine approaches the Carnot limit. 

That heat engines become ``weak'' in the Carnot limit, in that the power they deliver becomes vanishingly small, is a well known property. Here however, we presented a novel aspect, independent of time scales. We see that the Carnot limit is weak in that the weight must gain an infinite amount of energy before winning against its spread. Thus this is intrinsically a pathological point.

{

\section{Universality of efficiency of smallest thermal machines}\label{s:universality}
In the introduction we argued that the smallest thermal machines are the most fundamental thermal machines, and because of this their study shows the inner workings of thermodynamics in the clearest way. Here we will show the insight they provide into the efficiency of thermal machines.

The Carnot efficiency of a thermal machine is a universal property of the machine -- it depends only upon the temperature of the heat baths between which it works, but not upon the details of the construction of the machine or its interactions with the thermal bath. On the other hand, the efficiency of a thermal machine when operating away from the Carnot efficiency is a non-universal property, which does depend upon all of these details. In this section we will show however that 
\begin{quote}
\emph{the efficiency of the smallest machines is always universal -- both at the Carnot limit and away from it.}
\end{quote}
We shall focus here on the smallest refrigerator. Identical conclusions follow for the heat pump and engine. The idealised thermal machine which the smallest refrigerator is a specific instance of is depicted in FIG.~\ref{f:idealised} (a). The efficiency $\eta_{\text{fr}}$ for such a refrigerator is the ratio of the heat $Q_3$ extracted from the `cold' external bath at temperature $T_3$ to the heat $Q_1$ provided to the machine by the `hot' bath at temperature $T_1$,
\begin{equation}
	\eta_{\text{fr}} = \frac{Q_3}{Q_1}
\end{equation}
For the smallest machines it was shown previously \cite{Pop10b} that the ratio of the heat exchanged with each bath must match perfectly the ratio of the energy-level spacing of the corresponding qubits,
\begin{equation}\label{e:ratios}
	Q_1 : Q_2 : Q_3 = E_1 : E_2 : E_3
\end{equation}
which arises due to the fact that there is only a single way in which the baths can exchange heat. Therefore the efficiency $\eta^q_{\text{fr}}$ of the smallest refrigerator is given by
\begin{equation}
	\eta^q_{\text{fr}} = \frac{E_3}{E_1}
\end{equation}
Using the definition of $T_v$, we can express this result not in terms of the energies $E_1$ and $E_3$, but instead in terms of the virtual and bath temperatures. We find that
\begin{equation}\label{e:eta fridge}
	\eta^q_{\text{fr}} = \eta^C_{\text{fr}} \Bigg(1-\frac{\frac{1}{T_v}-\frac{1}{T_3}}{\frac{1}{T_v}-\frac{1}{T_2}}\Bigg)
\end{equation}
where 
\begin{equation}
	\eta^C_{\text{fr}} = \frac{\frac{1}{T_2}-\frac{1}{T_1}}{\frac{1}{T_3}-\frac{1}{T_2}}
\end{equation}
is the standard Carnot efficiency for a refrigerator driven by baths at $T_1$ and $T_2$ and extracting heat from a bath at temperature $T_3$ (see FIG.~\ref{f:idealised} (a)). In section \ref{s:carnot} we will discuss this result with respect to approaching Carnot efficiency.

In a similar fashion, we can derive the efficiencies $\eta^q_{\text{hp}}$ of the smallest heat pump and $\eta^q_{\text{he}}$ of the heat engine, given by
\begin{align}\label{e:eta pump}
\eta^q_{\text{hp}} &= \eta^C_{\text{hp}} \Bigg(1-\frac{\frac{1}{T_3}-\frac{1}{T_v}}{\frac{1}{T_1}-\frac{1}{T_v}}\Bigg) \\ 
\eta^q_{\text{he}} &= \eta^C_{\text{he}} \Bigg(1-\frac{-\frac{1}{T_v}}{\frac{1}{T_1}-\frac{1}{T_v}}\Bigg)\label{e:eta engine}
\end{align}
where
\begin{align}
	\eta^C_{\text{hp}} &= \frac{\frac{1}{T_1}-\frac{1}{T_2}}{\frac{1}{T_1}-\frac{1}{T_3}}& \eta^C_{\text{he}} &= 1-\frac{T_1}{T_2}
\end{align}
are the Carnot efficiency for a heat pump and heat engine driven by baths at $T_1$ and $T_2$, and delivering heat to a bath at $T_3$, or work, respectively (see FIG.~\ref{f:idealised} (b) - (c)).

Thus we see that for the smallest machines the efficiency is \emph{universal} -- it depends only upon the temperatures of two external baths and the single virtual temperature in the composite bath which the machine selects and couples to. It does not depend on any other details, such as the coupling constants between the machine and the bath, and the machine and the external system. 

The lack of universality of the efficiency of other thermal machines can now be better understood from this point of view. A general thermal machine will not select out a single virtual temperature, but rather a range of virtual temperatures, and couple an external system to all of them. The efficiency is therefore an average, which depends upon the relative coupling to each of these virtual temperatures, which is a machine-dependent and non-universal property. That is, it is precisely the ability to `perfectly filter' a single virtual temperature which gives the smallest machine its universal character.

\section{Reaching Carnot efficiency}\label{s:carnot}
Previously in our study of the smallest thermal machines we studied the question of whether they can approach the Carnot efficiency. Naively one may think that imposing restrictions upon the size of a machine may necessarily constrain it so that it cannot, even in principle, operate anywhere close to the Carnot efficiency. This however is not the case; as shown previously \cite{SkrBruLin10} the smallest machines can approach the Carnot efficiency. What we will show here  is a surprising twist 
\begin{quote}
	\emph{Essentially only the smallest machines can approach Carnot efficiency.}
\end{quote}
(a more detailed and qualified statement shall be presented below). 

In the previous section we showed that the efficiency of the smallest machines can be written as the Carnot efficiency plus a correction term, equations \eqref{e:eta fridge} -- \eqref{e:eta engine}. To approach the Carnot limit this correction term must vanish. Hence the machine must be fine-tuned to couple to the specific virtual temperature that makes this term vanish. For a machine constructed by two physical qubits, this may be achieved easily by tuning the energy level spacing $E_1$ and $E_2$ appropriately.

Let us consider now more general thermal machines. As stated previously, unlike the smallest thermal machines, which filter out a \emph{single} virtual temperature from the composite bath, generally a thermal machine will filter out a \emph{range} of virtual temperatures. The above analysis shows that if a machine filters out a temperature away from the precise one required, then it will not operate close to the Carnot limit. Thus \emph{in order to approach Carnot efficiency general thermal machines must be engineered to decouple from the un-wanted virtual temperatures.} i.e. the machine must filter out a range of virtual temperatures which approach the single desired temperature in the limit.

The only way to achieve this de-coupling is for the machine to be engineered in such a way that it effectively implements a smallest thermal machine. Indeed, consider a single pair of energy levels at spacing $E_3$ in the external system which the machine is acting upon. Given fixed external bath temperatures $T_1$ and $T_2$ we noted previously that there is only a \emph{single} choice of spacings $E_1$ and $E_2$ that both couple to $E_3$, (i.e. $E_2-E_1 = E_3$) and also produce a machine virtual qubit at temperature $T_v$. On the other hand, if the parts of the machine in contact with each bath have other spacings $E_1'$ and $E_2'$ which can couple to $E_3$ (i.e. $E_2' - E_1' = E_3$) then they lead to other virtual temperatures $T_v'$ different from the required $T_v$. Coupling the machine to these other virtual temperatures will spoil its efficiency. Thus we conclude that \emph{any thermal machine which filters a single virtual temperature must be coupled by only a single spacing to each bath and is hence exactly the smallest thermal machine or a collection thereof.} 

To summarise, quantum mechanics offers for free systems with discrete energy levels. All we need to do is ensure the spacing is the desired one. On the other hand, classical systems have essentially a continuous energy spectrum, so we cannot avoid them having undesired energy level spacings. It is nevertheless possible to approach Carnot efficiency if we engineer the machine in such a way that the undesirable spacings that couple to the system are all close to the desired virtual temperature. Hence classical machines that approach the Carnot limit do so by effectively removing energy levels from their spectrum, and becoming essentially identical to the smallest machines.\footnote{Note that we define the size of thermal machines by the dimension of their Hilbert space.} 

Finally, we would like to comment on the meaning of the Carnot limit, and the way in which reversibility is achieved. As it is well known, whenever two bodies at two different temperatures are put in contact, irreversibility occurs. The only way to achieve reversibility is to ensure that all systems that are in contact have the same temperature. This is precisely what happens in a machine which works in the reversible regime. As we can see from \eqref{e:eta fridge} and \eqref{e:eta pump}, for the refrigerator and heat pump the correction term vanishes and the Carnot limit is achieved precisely when the bodies which are placed into thermal contact -- the external system, the machine virtual qubit, and the selected virtual qubits of the composite bath -- all approach the same temperature. Put differently, one may be tempted to think that we have a situation in which three bodies at different temperatures are in contact with each other and this may necessarily lead to irreversibility. The difficulties that may arise from such a situation were exposed nicely by Parrondo \cite{ParEsp96}. We see however that reversibility can occur even in such a situation precisely because the relevant systems which interact in this situation are both at the same temperature.   

\section{Entropy}\label{s:entropy}
Analysing the flow of entropy in thermal machines is one of the most basic aspects of thermodynamics. Here we will do so for the smallest thermal machines, where we find that it is highly illuminating. In particular we will see that entropically as well the virtual qubit behaves as a real qubit at the virtual temperature, reinforcing our argument that the virtual temperature acts as a real temperature in all respects.
\begin{figure}
	\includegraphics[width=\columnwidth]{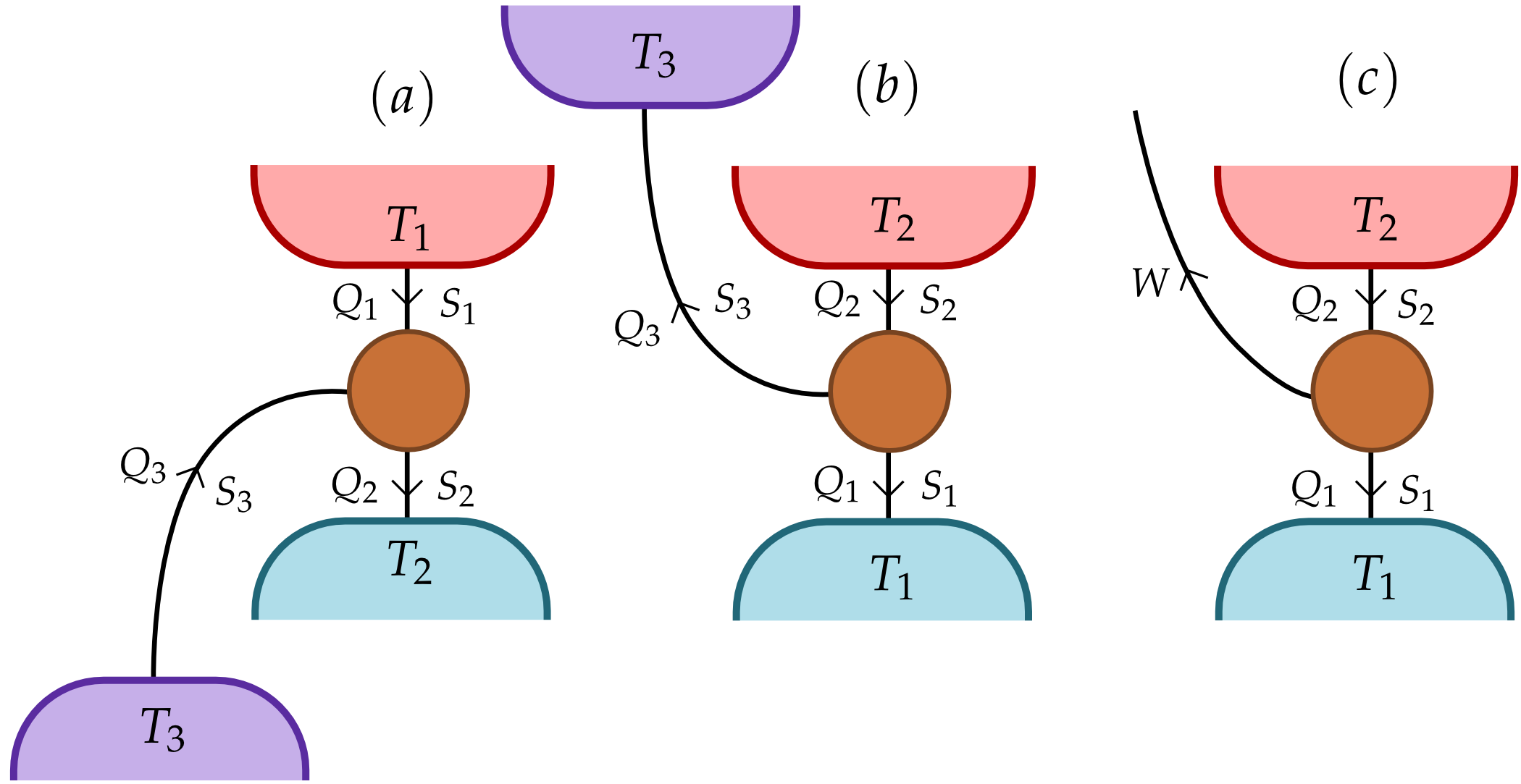} \caption{\label{f:idealised} (Colour online) {\bf Schematic diagram of idealised thermal machines:} Schematic diagram, showing the flow of heat and entropy for (a) refrigerator (b) heat pump (c) heat engine.}
\end{figure}
Let us focus on the smallest refrigerator, depicted in idealised form in FIG.~\ref{f:idealised}. As noted previously, the smallest machines have the property that the ratio of heat currents match the ratio of energy levels, \eqref{e:ratios}. Using this fact, in conjunction with the definition of $T_v$, \eqref{e:Tv machine}, we obtain
\begin{equation}\label{e:entropy eq}
	\frac{Q_2}{T_2} - \frac{Q_1}{T_1} = \frac{Q_3}{T_v}
\end{equation}
which is clearly an entropy equality involving the entropies $S_1 = Q_1/T_1$, $S_2 = Q_2/T_2$, flowing out of bath 1 and into bath 2, and the entropic quantity $S_v = Q_3/T_v$.

The change in total entropy of the two baths and external system for the refrigerator is given by
\begin{equation}
	\Delta S_{\text{fr}} = S_2 - S_1 - S_3
\end{equation}
where $S_3 = Q_3/T_3$ is the entropy flowing out of bath 3. Using  \eqref{e:entropy eq} this can be re-written as
\begin{equation}\label{e:entropy}
	\Delta S_\text{fr} = \frac{Q_3}{T_v} - \frac{Q_3}{T_3}
\end{equation}
By similar analysis, for the heat pump and heat engine we find
\begin{align}
	\Delta S_\text{hp} &= \frac{Q_3}{T_3}-\frac{Q_3}{T_v}& \Delta S_\text{he} &= -\frac{Q_3}{T_v}
\end{align}

The meaning of these results (for the refrigerator and heat pump) are as follows: We have two bodies in thermal contact -- the external system at temperature $T_3$, and the virtual qubits in the composite bath at temperature $T_v$. The bodies exchange the quantity of heat $Q_3$. Now we see that the meaning of the entropic quantity $Q_3/T_v$ is the entropy of the virtual qubits, and the relation \eqref{e:entropy} is the standard entropy flow for two bodies in thermal contact.

Similar analysis of a machine putting heat into a body at $T_v <0$ shows how the strength of work, as characterised by the smallness of the absolute value of $T_v$,
is related to the amount of entropy that is removed from the body while delivering a fixed amount of heat $Q_3$: the smaller the absolute value of $T_v$ the larger
the entropy removed from the body, with the amount tending to infinity as $T_v \to 0$ from below.

Incidentally, this shows again that the Carnot limit is approached exactly when the two systems in thermal contact approach the same temperature. 
\section{Genuine thermal machines}\label{s:genuine}

We would like to define now an important concept, that of \emph{genuine refrigerators and heat pumps}. Your
refrigerator at home uses a supply of work to transfer heat from a cold source to a hot source. Furthermore,
from a theoretical point of view, the way refrigerators are generally presented is that of ``an engine running
in reverse'', using a supply or work to move heat against the gradient. The point we want to make now is that
\emph{work is not necessary}.

On the one hand it is well known that an \emph{external} source of work is not required to produce a
refrigerator. However, by itself this fact is not surprising -- we can always replace the source of work with an
engine that produces it, using two external baths. Indeed, we know that the work used to power our refrigerator
at home ultimately comes from a power station. There are also examples of refrigerators in which there is no
apparent place in which work is used - adsorption refrigerators for example have no engines and pumps -- but this
by itself doesn't mean that work is not present in some hidden form (such as while gases are expanding, etc.)
The concepts of virtual qubits and virtual temperatures offer for the first time the tools to address this
question.

The crucial point we want to make is that it is not necessary to generate work -- even internally in the machine
-- to create a refrigerator.  The refrigerators we presented in the previous section function by putting an
external system in contact with virtual qubits whose temperature is lower than $T_1$ and $T_2$, the actually
temperatures of the two baths. At no point is there ever a population inversion involved -- in other words at no
point is work produced or used.

In this sense, this is a \emph{genuine} refrigerator. Ordinary refrigerators (and heat pumps) that use work are
now seen to be ``wasteful''. This is most evident if we talk about the \emph{resources} that a composite thermal
bath provide us with. It is clear that work is a resource provided by the composite bath, and we can indeed use
this to produce a refrigerator. However, we see that the composite bath provides other resources as well, namely
\emph{energy at any temperature}, not merely work. Energy at a cold temperature is a resource which can be used
to achieve cooling directly, while work needs to be somehow converted. Thus ordinary refrigerators make
non-necessary use of a resource (work) that is more powerful than what is needed. Genuine thermal machines use
the minimal possible resources, and are thus the ``purest'' thermal machines.

Furthermore, it is essential to make the difference between genuine and non-genuine thermal machines when we
want to know what exactly do they do to the thermal baths that drive them. If we take the baths not to be
strictly infinite in size, then thermal machines necessarily degrade them, in particular reducing their free
energy. Genuine refrigerators couple to different virtual qubits than refrigerators which use work as an
intermediate effect. Hence they clearly affect the bath in a different way. It would be interesting to study
further exactly these differences.

Finally, it is illuminating to see, for the case of the smallest machines, how the functionality is changed as we vary the bath parameters. In FIG.~\ref{f:machines} we hold fixed the design of the machine ($E_1$ and $E_2$), and the bath temperature $T_1$, and plot how the virtual temperature (and therefore function) changes as we vary  the remaining bath temperature $T_2$. It is interesting to note that as $T_2$ approaches $\tfrac{E_2}{E_1}T_1$ from above then $T_v \to \infty$. This engine is the reversible (Carnot) engine, and is seen to be at the transition  between heat pump and heat engine; in this sense it is again seen to be the `weak'.
\begin{figure}[h]
    \includegraphics[width=0.9\columnwidth]{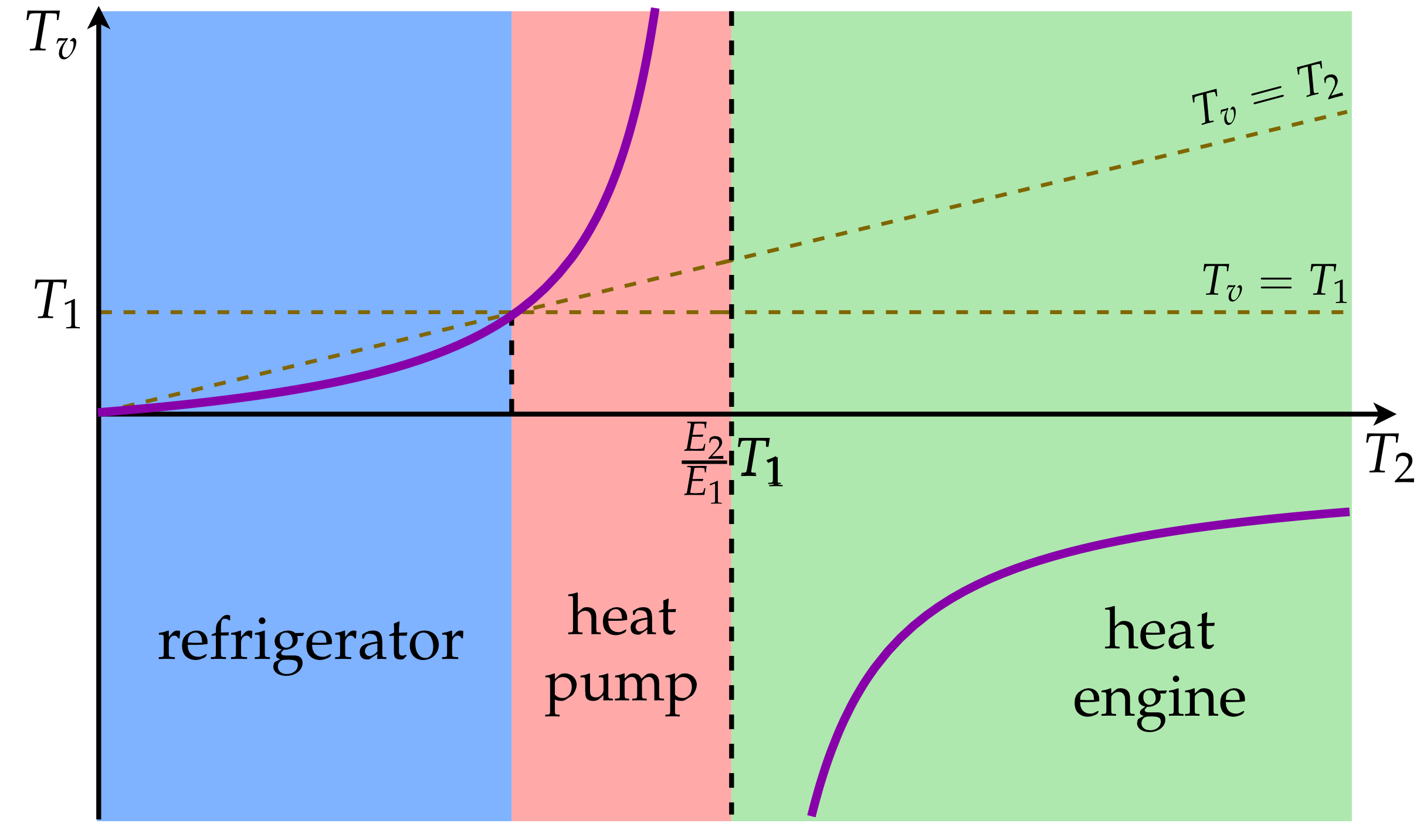} \caption{\label{f:machines} (Colour online) {\bf Graph of virtual temperature against bath temperature:} We hold fixed the local energy-level spacings $E_1$ and $E_2$ as well as the bath temperature $T_1$, and allow the bath temperature $T_2$ to vary. When $T_2 < T_1$, the virtual temperature $T_v$ becomes smaller than either environmental temperature and the machine is a refrigerator. For $T_1 < T_2 < \tfrac{E_2}{E_1}T_1$, $T_v$ becomes larger than $T_1$ and $T_2$ and hence the machine is a heat pump. Finally, for $\tfrac{E_2}{E_1}T_1 < T_2$, $T_v$ becomes negative and the machine functions as a heat engine. }
\end{figure}

{\bf Acknowledgements} The authors acknowledge support from the UK EPSRC and the European project Q-ESSENCE.


\begin{appendix}

\section{Thermal machines acting on finite dimensional systems}\label{a:bounded} 
In this appendix we will demonstrate explicitly our claim, that external systems put into thermal contact with the machine virtual qubit thermalise to its virtual temperature, for the case of finite dimensional systems. Crucially this holds for both positive \emph{and} negative temperatures. The temperature which the external system thermalises to classifies the behaviour of the machine: If it is colder than either environmental temperature then the machine is a refrigerator; if it is hotter then it is a heat pump. Finally, if the temperature is negative, then it is a heat engine.

To see explicitly this thermalisation, consider initially that the external system is itself completely isolated except for the interaction with the thermal machine.  That is, we consider the external system to only be in contact with the thermal machine, not with any other external system, so it is only the machine that determines its behaviour.

In the weak coupling limit, the dynamics of the system are accurately described by a master equation. The equation will generically take the form 
\begin{equation}\label{e:master qubit} 
	\frac{\partial \rho}{\partial t} = -i[H_0+H_{int},\rho] + \mathcal{D}_1(\rho) + \mathcal{D}_2(\rho) 
\end{equation}
where
\begin{equation}
    H_0=E_1\proj{1}{1}+E_2\proj{1}{2} +E_3\proj{1}{3}
\end{equation}
\begin{multline}\label{e:Hint appendix}
	H_{int} = g\Big(\kets{0}{1}\kets{1}{2}\kets{0}{3}\bras{1}{1}\bras{0}{2}\bras{1}{3} \\+ \kets{1}{1}\kets{0}{2}\kets{1}{3}\bras{0}{1}\bras{1}{2}\bras{0}{3}\Big)
\end{multline}
with $E_3 = E_2-E_1$, and $\mathcal{D}_1(\rho)$ and $\mathcal{D}_2(\rho)$ are the dissipative dynamics acting on qubit 1 and 2 respectively. Such an equation provides a consistent description of the dynamics of the system when both the strength of the interaction Hamiltonian and the dissipative dynamics are weak \cite{Tal86}.  

Without specifying a specific model for the dissipative dynamics, all such models must satisfy 
\begin{align}\label{e:dissipative} 
	\mathcal{D}_1(\tau_1\otimes\sigma_{23}) &= 0& \mathcal{D}_2(\tau_2\otimes\sigma_{13}) &= 0 
\end{align}
where 
\begin{align}
	\tau_i &= \tfrac{1}{\mathcal{Z}_i}e^{-H_i/T_i}& \mathcal{Z}_i &= \tr(e^{-H_i/T_i})
\end{align} 
are the thermal equilibrium state and partition function of qubit $i$, and the states $\sigma_{23}$ and $\sigma_{13}$ are arbitrary density matrices. In other words, a necessary requirement for any weakly coupled and weakly interacting system is that the thermal equilibrium state of each qubit is the stationary state of its dissipative dynamics.
	
To find the stationary solution to \eqref{e:master qubit} it suffices to note that 
\begin{multline}\label{e:Ht qubit} 
	\left[\tfrac{H_1}{T_1} + \tfrac{H_2}{T_2} + \tfrac{H_3}{T_3},H_0 + H_{int}\right] = 
		g\left(\tfrac{E_2}{T_2} - \tfrac{E_1}{T_1} -\tfrac{E_3}{T_3}\right)\\ \times \Big(\kets{0}{1}\kets{1}{2}\kets{0}{3}\bras{1}{1}\bras{0}{2}\bras{1}{3} - \kets{1}{1}\kets{0}{2}\kets{1}{3}\bras{0}{1}\bras{1}{2}\bras{0}{3}\Big)
\end{multline}
Thus, by recalling the definition of the virtual temperature given in the main text, equation \eqref{e:Tv machine}, it is evident that it is exactly when $T_3 = T_v$ that the right-hand-side of \eqref{e:Ht qubit} vanishes. In this instance the operator appearing on the left of the commutator commutes with the total (free + interaction) Hamiltonian of of the system. If we thus define the state 
\begin{equation}
	\tau_v = \tfrac{1}{\mathcal{Z}_v} e^{-H_3/T_v} 
\end{equation}
which is simply the thermal equilibrium state of qubit 3 at temperature $T_v$, then it follows immediately that the thermal product state 
\begin{equation}
		\tau_1 \otimes \tau_2 \otimes\tau_v \equiv\tfrac{1}{\mathcal{Z}_1}\tfrac{1}{\mathcal{Z}_2}\tfrac{1}{\mathcal{Z}_v}\exp\left(-\tfrac{H_1}{T_1}-\tfrac{H_2}{T_2}-\tfrac{H_3}{T_v}\right) 
\end{equation}
also commutes with the total Hamiltonian -- showing that it is stationary with respect to the unitary dynamics. This state is however also seen to satisfy conditions \eqref{e:dissipative}, showing that it is stationary with respect to the dissipative dynamics. Combining these two facts we thus conclude that it is the stationary solution of the master equation \eqref{e:master qubit}.
	
Thus the external qubit will approach a thermal state at the temperature of the machine virtual, qubit showing that it indeed thermalises \emph{to the temperature of the virtual qubit}, as stated. 

It is important to realise that this result can be extended beyond the thermalisation of qubits. Indeed, qubits have the special property that every diagonal density matrix (in the energy eigenbasis) can have a temperature associated to it. Thus to show that the external system is indeed \emph{thermalising} to the temperature of the virtual qubit we must consider more general systems. Let us consider therefore an $N$-level system which has equidistance spacing between its energy eigenstates. That is, we consider a system with energy eigenstates $\ket{0},\ldots,\ket{N-1}$, and free Hamiltonian 
\begin{equation}\label{e:H3[N]} 
	H_3^{[N]} = \sum_{n=0}^{N-1} n E_3 \proj{n}{3}
\end{equation}
where we maintain the condition that $E_3 = E_2-E_1$. This system is taken to interact with the two machine qubits via the interaction Hamiltonian 
\begin{multline}\label{e:Hint quNit} 
	H_{int}^{[N]} = g\sum_{n=0}^{N-2}\kets{0}{1}\kets{1}{2}\kets{n}{3}\bras{1}{1}\bras{0}{2}\bras{n+1}{3} \\+ \kets{1}{1}\kets{0}{2}\kets{n+1}{3}\bras{0}{1}\bras{1}{2}\bras{n}{3}
\end{multline}
which can be seen as a sum of terms, all of the form \eqref{e:Hint appendix}. Using an identical proof method as used for qubits, or by direct substitution, it is straightforward to show that the thermal product state 
\begin{equation}
	\tau_1 \otimes \tau_2 \otimes\tau_v^{[N]} \equiv \tfrac{1}{\mathcal{Z}_1\mathcal{Z}_2\mathcal{Z}_v^{[N]}}\exp\left(-\tfrac{H_1}{T_1}-\tfrac{H_2}{T_2}-\tfrac{H_3^{[N]}}{T_v}\right) 
\end{equation}
is the stationary solution to the corresponding master equation, where 
\begin{align}
	\tau_v^{[N]} &= \tfrac{1}{\mathcal{Z}_v^{[N]}}e^{-H_3^{[N]}/T_v}& \mathcal{Z}_v^{[N]} &= \tr\left(e^{-H_3^{[N]}/T_v}\right) 
\end{align}
In other words, \emph{all equi-spaced systems, when placed into thermal contact with the virtual qubit (via the interaction \eqref{e:Hint quNit}) thermalise to the virtual temperature}. Crucially, when the virtual temperature is negative, the final state is an \emph{inverse Boltzmannian}, with population decreasing exponentially with decreasing energy as we move away from the \emph{most excited state} $\ket{N-1}$.
\begin{figure}
	\includegraphics[height=40mm]{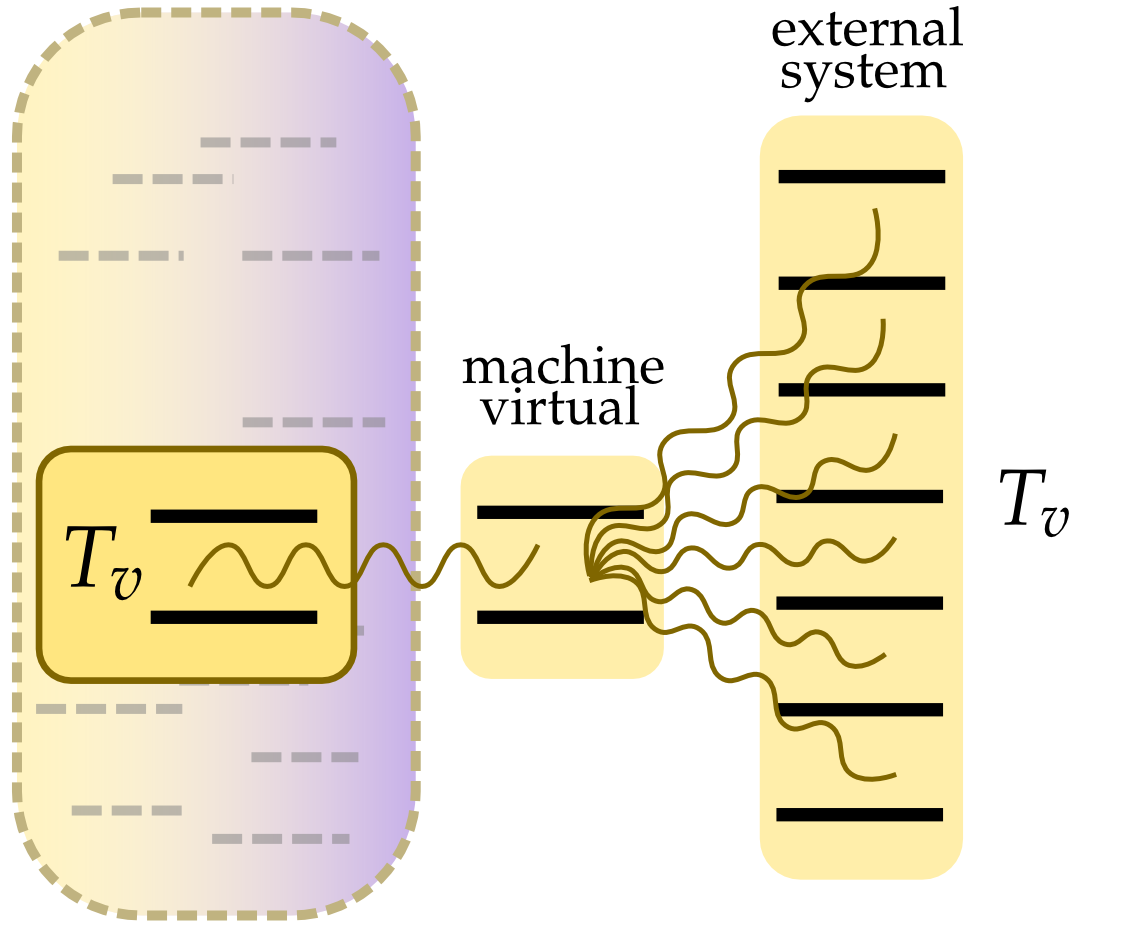} \caption{\label{f:small-machine-N} (Colour online) {\bf Schematic diagram of smallest machine interacting with an isolated external system:} An isolated external system -- here an $N$ level equi-spaced system -- is placed into thermal contact (wavey lines) with the machine virtual qubit, which has matching energy-level spacing. The net effect is that the external system is placed in thermal contact with virtual qubits in the composite bath at temperature $T_v$, mediated by the machine virtual qubit, and reaches a Boltzmannian at temperature $T_v$. This holds independent of whether $T_v$ is positive, or negative, in which case the state is an `inverted Boltzmannian'.}
\end{figure}
	
If instead of considering the external system to be isolated we place it in contact with a third thermal bath at some fixed temperature $T_3$ then we would expect to find a non-universal steady-state solution to \eqref{e:master qubit} which will depend upon the explicit form the of dissipative terms $\mathcal{D}_i(\rho)$. 

In \cite{LinPopSkr10a}, for the case where the external object is a qubit (to be cooled), a specific model was employed and it was shown that with three thermal baths the external object reaches a stationary state which is cooler than its environment. From the current perspective the situation is simple to understand; the object is in thermal contact with both the virtual qubit and also its own environment. We thus expect it to reach a steady state with temperature between these two temperatures, with the precise temperature depending on the relative `magnitude' of the two thermalising effects. Therefore as long as $T_v < T_3$ intuitively we expect the object to be cooled, which is exactly what was found for the model presented in \cite{LinPopSkr10a}.

\section{Thermal machines acting on infinite dimensional systems}\label{a:heat engine} 
In this appendix we will present the detailed calculations of a specific model which shows that when we take the external system to be unbounded in energy, and engineer the smallest thermal machine to have the virtual qubit at a negative virtual temperature, then the system is continually excited, and thus captures the classical notion of a heat engine producing work, as claimed in the main text.

As explained earlier, to do so we place the machine virtual qubit in thermal contact with a ``weight'' -- a system consisting of equally-spaced energy levels, unbounded from above and below. Namely we have the free and interaction Hamiltonians
\begin{equation} 
	H_0=E_1\proj{1}{1}+E_2\proj{1}{2} + \sum_{n=-\infty}^{\infty} nE_w\proj{n}{w}
\end{equation}
\begin{multline}
	H_{int}= g\sum_{n=-\infty}^{\infty}\kets{0}{1}\kets{1}{2}\kets{n}{w}\bras{1}{1}\bras{0}{2}\bras{n+1}{w} \\+ \kets{1}{1}\kets{0}{2}\kets{n+1}{w}\bras{0}{1}\bras{1}{2}\bras{n}{w}
\end{multline}
where $E_1$ and $E_2$ are chosen such that $E_2-E_1 = E_w$. To complete the model we must give an explicit form for the dissipative dynamics of the machine, i.e. a model for the thermalisation of each machine qubit with its bath. We will use the same model used in our previous work on the smallest refrigerator \cite{LinPopSkr10a}, namely
\begin{equation}
	\mathcal{D}_i(\rho) = p(\tau_i \otimes \tr\rho - \rho)
\end{equation}
This simple model of thermalisation clearly satisfies the conditions \eqref{e:dissipative} and and generates an exponential decay of each qubit back to its thermal state $\tau_i$, with decay constant $p$. The explicit master equation which governs the dynamics of the system therefore takes the form 
\begin{equation}\label{eqn-of-motion1} 
	\frac{ \partial \rho}{\partial t} = -i[H_0+H_{int}, \rho] + \sum_{i=1}^2 p(\tau_i \tr_i \rho - \rho) 
\end{equation}
We expect that this equation accurately models the behaviour of the system as long as $g$ and $p$ are chosen such that $g, p \ll E_k$, so that we are in the weak-dissipation weak-interaction regime. Here it is consistent both to use the free Hamiltonian to define the thermal states, and to neglect additional contributions to the dissipative dynamics originating from the interaction between the qubits. Note that here we make the simplifying assumption that the two baths interact with the same coupling strength $p$. The analysis presented below can also be carried out in the case of unequal coupling strengths $p_1 \neq p_2$, however it substantially complicates the algebra without changing the results obtained. 
\subsection{Raising of the weight}	
By first multiplying \eqref{eqn-of-motion1} by the free Hamiltonian of the weight $H_{w}$, and then taking the trace, we find an expression for the rate of change of the average energy of the weight, given by 
\begin{equation}\label{e:ExpEw} 
	\frac{d}{dt}\Exp{E_w} = \frac{d}{dt} \tr(H_w\rho)= -igE_w\Delta 
\end{equation}
where 
\begin{multline}
	\Delta = \sum_n \bras{0}{1}\bras{1}{2}\bras{n}{w}\rho\kets{1}{1}\kets{0}{2}\kets{n+1}{w} \\ - \bras{1}{1}\bras{0}{2}\bras{n+1}{w}\rho\kets{0}{1}\kets{1}{2}\kets{n}{w}
\end{multline}
By introducing the three new operators on qubits 1 and 2, 
\begin{eqnarray}
	Z_v &=& \kets{1}{1}\kets{0}{2}\bras{1}{1}\bras{0}{2}-\kets{0}{1}\kets{1}{2}\bras{0}{1}\bras{1}{2} \nonumber \\
	\overline{Z_v} &=& \kets{0}{1}\kets{0}{2}\bras{0}{1}\bras{0}{2}-\kets{1}{1}\kets{1}{2}\bras{1}{1}\bras{1}{2} \\
	N_v &=& \kets{1}{1}\kets{0}{2}\bras{1}{1}\bras{0}{2}+\kets{0}{1}\kets{1}{2}\bras{0}{1}\bras{1}{2} \nonumber 
\end{eqnarray}
	and denoting $\Exp{Z_v} = \tr\big(\rho Z_v\big)$, $\Exp{\overline{Z_v}} = \tr\big(\rho \overline{Z_v}\big)$ and $\Exp{N_v} = \tr\big(\rho N_v\big)$ then the following set of equations can be obtained in a straightforward manner from \eqref{eqn-of-motion1} 
	\begin{eqnarray}
		\frac{d}{dt}\Delta &=& - 2p\Delta \nonumber -2ig\Exp{Z_v} \\
		\frac{d}{dt} \Exp{Z_v} &=& -2ig\Delta - p\Big(\Exp{Z_v}-\Exp{Z_v^{eq}}\Big) \\
		\frac{d}{dt}\Exp{N_v} &=& p\Big(1-2\Exp{N_v} +\Exp{Z_v^{eq}}\Exp{Z_v}-\Exp{\overline{Z_v}^{eq}}\Exp{\overline{Z_v}}\Big) \nonumber \\
		\frac{d}{dt} \Exp{\overline{Z_v}} &=& - p\Big(\Exp{\overline{Z_v}}-\Exp{\overline{Z_v}^{eq}}\Big) \nonumber 
	\end{eqnarray}
	where 
	\begin{align}
		\Exp{Z_v^{eq}} &= \tr\big(Z_v\tau_1\otimes\tau_2\big)\nonumber\\
		\Exp{\overline{Z_v}^{eq}} &= \tr\big(\overline{Z_v}\tau_1\otimes\tau_2\big) 
	\end{align}
	are the thermal equilibrium expectation values of the operators $Z_v$ and $\overline{Z_v}$. We see that the evolution of $\Exp{\overline{Z_v}}$ is completely independent of the evolution of the other variables -- a situation which only occurs when the two baths have a common coupling strength $p$. Furthermore, although the evolution of $\Exp{N_v}$ depends on all the other variables, we see that no other variable depends directly upon it. This system of equations can easily be solved for the asymptotic (stationary) behaviour. Denoting the variables in this limit with a superscript `S' we find that 
\begin{align}\label{e:asymptotic solution I} 
	\Delta^S &= -\frac{igp}{2g^2+p^2}\Exp{Z_v^{eq}} \nonumber \\
	\Exp{Z_v^S} &= \frac{p^2}{2g^2+p^2} \Exp{Z_v^{eq}} \nonumber \\
	\Exp{N_v^S} &= \Exp{N_v^{eq}} - \frac{g^2}{2g^2+p^2}\Exp{Z_v^{eq}}^2 \\
	\Exp{\overline{Z_v}^S} &= \Exp{\overline{Z_v}^{eq}} \nonumber 
\end{align}
There are a number of notable features of this solution. The first is to note that $\Exp{\overline{Z_v}^S}$ is the `bias' of the ``anti-virtual'' qubit -- the virtual qubit formed from the states $\ket{00}$ and $\ket{11}$, and that this bias is left unaltered by the coupling of the virtual qubit to the weight. Second, we see that the normalisation of the virtual qubit is shifted from its value at thermal equilibrium and that its stationary value is only ever smaller than (or equal to) this value. Finally, we see that the bias of the virtual qubit is scaled by the coupling, and always decreases in magnitude.

Finally, but substitution of the solution \eqref{e:asymptotic solution I} into \eqref{e:ExpEw}, we obtain for the asymptotic rate of change of the average energy
\begin{equation}\label{e:rate}
    \frac{d}{dt}\langle E_w\rangle = -\frac{g^2p}{2g^2+p^2}E_w \langle Z_v^{eq} \rangle
\end{equation}
\subsection{Spreading of the weight}	
	A second quantity of interest is $\Exp{E_w^2}$, which gives information about the spreading of the average energy of the weight in time. From \eqref{eqn-of-motion1} it follows that the equation governing the evolution of this quantity is 
\begin{equation}
	\frac{d}{dt}\Exp{E_w^2} = -igE_w^2\Delta_n 
\end{equation}
where we have introduced the quantity $\Delta_n$, defined by 
\begin{multline}
	\Delta_n = \sum_{n=-\infty}^{\infty} (2n+1)\Big(\bras{0}{1}\bras{1}{2}\bras{n}{w}\rho\kets{1}{1}\kets{0}{2}\kets{n+1}{w}\\-\bras{1}{1}\bras{0}{2}\bras{n+1}{w}\rho\kets{0}{1}\kets{1}{2}\kets{n}{w}\Big) 
\end{multline}
Introducing a second new quantity $\Sigma_n$, related in an analogous manner to $\Exp{Z_v^S}$ as $\Delta_n$ is to $\Delta$, 
\begin{multline}
	\Sigma_n = \sum_{n=-\infty}^{\infty} (2n+1)\Big(\bras{1}{1}\bras{0}{2}\bras{n+1}{w}\rho\kets{1}{1}\kets{0}{2}\kets{n+1}{w}\\-\bras{0}{1}\bras{1}{2}\bras{n}{w}\rho\kets{0}{1}\kets{1}{2}\kets{n}{w}\Big) 
\end{multline}
Then it is possible to obtain the following set of coupled differential equations 
\begin{align}
	\frac{d}{dt}\Delta_n &= -2p \Delta_n -2ig\Sigma_n \\
	\frac{d}{dt}\Sigma_n &= -2ig\Delta_n - p \Sigma_n -2g^2t\Exp{Z_v^{eq}}\Exp{Z_v^S}-p\Exp{N_v^S}\nonumber 
\end{align}
which are valid in the asymptotic limit, since to derive the second equation the asymptotic solution \eqref{e:asymptotic solution I} has been used. These equations can easily be solved using standard techniques and the asymptotic solution for the expected-squared-energy is given by 
\begin{multline}\label{e:rate sq}
	\frac{d}{dt}\Exp{E_w^2} = \frac{d}{dt}\Exp{E_w}^2 +\frac{g^2p}{2g^2+p^2}E_w^2\Big(\Exp{N_v^{eq}}\\ -\frac{2g^2(g^2+2p^2)}{(2g^2+p^2)^2}\Exp{Z_v^{eq}}^2\Big)
\end{multline}
\subsection{Heat transfers}
Finally, if we define further the two quantities $\Gamma_1$ and $\Gamma_2$, which are the instantaneous ground state probabilities for qubits 1 and 2 respectively, 
\begin{multline}
		\Gamma_1 = \sum_n\bras{0}{1}\bras{0}{2}\bras{n}{w}\rho\kets{0}{1}\kets{0}{2}\kets{n}{w} \\ + \bras{0}{1}\bras{1}{2}\bras{n}{w}\rho\kets{0}{1}\kets{1}{2}\kets{n}{w}
\end{multline}
\begin{multline}
		\Gamma_2 = \sum_n\bras{0}{1}\bras{0}{2}\bras{n}{w}\rho\kets{0}{1}\kets{0}{2}\kets{n}{w} \\ + \bras{1}{1}\bras{0}{2}\bras{n}{w}\rho\kets{1}{1}\kets{0}{2}\kets{n}{w}
\end{multline}
then these quantities obey the coupled set of equations 
\begin{eqnarray}
	\frac{d}{dt} \Gamma_1 &=& +ig\Delta + p\big(\Gamma_1^{eq} - \Gamma_1\big) \\
	\frac{d}{dt}\Gamma_2 &=& -ig\Delta + p\big(\Gamma_2^{eq} - \Gamma_2\big) \nonumber 
\end{eqnarray}
where $\Gamma_i^{eq} = (1+e^{-E_i/T_i})^{-1}$ is the equilibrium ground state population of each qubit. Making use of \eqref{e:asymptotic solution I} it can therefore be seen that asymptotically these populations reach the values 
\begin{align}
	\Gamma_1^S &= \Gamma_1^{eq} + \tfrac{g^2}{2g^2+p^2} \Exp{Z_v^{eq}} \nonumber \\
	\Gamma_2^S &= \Gamma_2^{eq} - \tfrac{g^2}{2g^2+p^2} \Exp{Z_v^{eq}} 
\end{align}
The rate at which heat flows between the qubits and their environments is given by the change in energy of each qubit due to the interaction with the baths. Given the master equation \eqref{eqn-of-motion1}, the asymptotic heat currents are therefore 
\begin{align}\label{e:heat} 
	\frac{d}{dt}Q_i &= \tr\big(H_i\mathcal{D}_i(\rho)\big)=p\tr\Big(H_i\big(\tau_i-\rho_i^S\big)\Big) \nonumber \\
	&= (-1)^{i+1}\frac{g^2p}{2g^2+p^2}E_i{\Exp{Z_v^{eq}}} 
\end{align}
\end{appendix}
\end{document}